\documentclass[aip, apl, reprint, 10pt, twocolumn, showpacs, preprintnumbers,superscriptaddress, amsmath, amssymb]{revtex4-2}
\usepackage[pdftex]{graphicx}
\usepackage[ngerman,english]{babel}
\usepackage{natbib}
\usepackage{color}
\usepackage{siunitx}
\usepackage{calc}
\usepackage[colorlinks=true, pdfstartview=FitV, linkcolor=blue,citecolor=blue, urlcolor=blue]{hyperref}
\usepackage{float}

\definecolor{green2}{rgb}{.0, .58, 0}

\begin{document}
\begin{abstract}
We use sensitive dynamic cantilever magnetometry to measure the magnetic hysteresis of individual magnetic Janus particles.  These particles consist of hemispherical caps of magnetic material deposited on micrometer-scale silica spheres.  The measurements, combined with corresponding micromagnetic simulations, reveal the magnetic configurations present in these individual curved magnets. In remanence, ferromagnetic Janus particles are found to host a global vortex state with vanishing magnetic moment.  In contrast, a remanent onion state with significant moment is recovered by imposing an exchange bias to the system via an additional antiferromagnetic layer in the cap.  A robust remanent magnetic moment is crucial for most applications of magnetic Janus particles, in which an external magnetic field actuates their motion.
\end{abstract}
\title{Magnetic hysteresis of individual Janus particles with hemispherical exchange biased caps}
\author{S.~Philipp} \affiliation{Department of Physics, University of Basel, 4056 Basel, Switzerland}
\author{B.~Gross} \affiliation{Department of Physics, University of Basel, 4056 Basel, Switzerland}
\author{M.~Reginka}\affiliation{Institute of Physics, University of Kassel, 34132 Kassel, Germany}
\author{M.~Merkel}\affiliation{Institute of Physics, University of Kassel, 34132 Kassel, Germany}
\author{M.~Claus} \affiliation{Department of Physics, University of Basel, 4056 Basel, Switzerland}
\author{M.~Sulliger} \affiliation{Department of Physics, University of Basel, 4056 Basel, Switzerland}
\author{A.~Ehresmann}\affiliation{Institute of Physics, University of Kassel, 34132 Kassel, Germany}
\author{M.~Poggio} \affiliation{Department of Physics, University of  Basel, 4056 Basel, Switzerland} \email{martino.poggio@unibas.ch}
\homepage{http://poggiolab.unibas.ch/}
\maketitle
\newpage
%
Janus particles (JPs) are nano- or micronsized particles that possess two sides, each having different physical or chemical properties. There is a multitude of types of JPs~\cite{walther_janus_2013}, differing in shape, material, and functionalization. 
As a subgroup of micron and sub-micron sized magnetic particles, which are discussed as a multi-functional component in lab-on-chip or micro-total analysis systems \cite{Issadore_2014_MagneticSensing, Ehresmann_2015_Superparamangetic}, magnetic JPs, consisting of a hemispherical cap of magnetic material on a non-magnetic spherical template, allow not only a controlled transversal motion, but also a controlled rotation by rotating external magnetic fields~\cite{McNaughton_2006_SuddenBreakdown, McNaughton_2011_Experimental}.
Such JPs can be mass-produced via the deposition of magnetic layers on an ensemble of silica spheres.
The transversal and rotary motion of these particles can be controlled via external magnetic fields, which exert magnetic forces and
torques~\cite{erb_actuating_2016}. This ability to externally actuate magnetic JPs has led to applications in microfluidics, e.g.\ as
stirring devices~\cite{steinbach_anisotropy_2020}, as microprobes for
viscosity changes~\cite{nguyen_detecting_2014}, or as cargo
transporters in lab-on-chip devices~\cite{baraban_transport_2012,
  valadares_catalytic_2010, golestanian_propulsion_2005}.  Magnetic
JPs have also been proposed as an in vivo drug delivery
system~\cite{bozuyuk_shape_2021}.

Although, in general, a transversal controlled motion can be achieved by both superparamagnetic particles or particles with a permanent magnetic moment, a control over the rotational degrees of freedom can only be achieved, if the particles possess a sufficiently large permanent magnetic moment.  Streubel et al.~\cite{streubel_equilibrium_2012} analyzed the remanent magnetic
state of magnetic JPs with ferromagnetic (fm) magnetic caps.  Their
simulations show that permalloy JPs with diameters larger than
$\SI{140}{\nano\meter}$ host a global vortex state at remanence.
Because this flux-closed state has a vanishing net magnetic moment,
magnetic JPs hosting such a remanent configuration are unsuited for
applications involving magnetic actuation.  Thus, for JPs larger than
this critical diameter, strategies to overcome this limitation need to
be developed.

Here, we make use of exchange bias~\cite{meiklejohn_new_1957,
  nogues_exchange_1999, Stamps_Exchange_2001}, which, in a simplified
picture imposes a preferred direction on the magnetic moments of the
fm layer and is thereby able to prevent the formation of a global
vortex at remanence.  We apply an exchange bias to the fm layer by
adding an antiferromagnetic (afm) layer beneath the fm layer.  In
order to verify that this addition leads to a remanent configuration
with large magnetic moment, we measure the magnetic hysteresis of
individual JPs with and without this layer.  The measurement of
individual JPs is necessary in order to eliminate the effects of
interactions between neighboring JPs.  For this task, we employ
dynamic cantilever magnetometry (DCM) and analyze the results by
comparison to corresponding micromagnetic simulations.  This technique
overcomes the limitation of earlier measurements, that were restricted
to ensembles of interacting JPs on a
substrate~\cite{tomita_magnetic_2021}.  These measurements relied on
the longitudinal magneto-optical Kerr effect and magnetic force
microscopy.  They found an onion state with a large remanent
magnetization in JPs with an afm layer.  Nevertheless, given that the
measurements were done on close-packed ensembles of JPs, they do not
exclude effects due to the interaction between the particles and,
therefore, cannot be used to infer the behavior of isolated JPs.\\

%
\begin{figure}[t]
  \includegraphics[width=.48\textwidth]{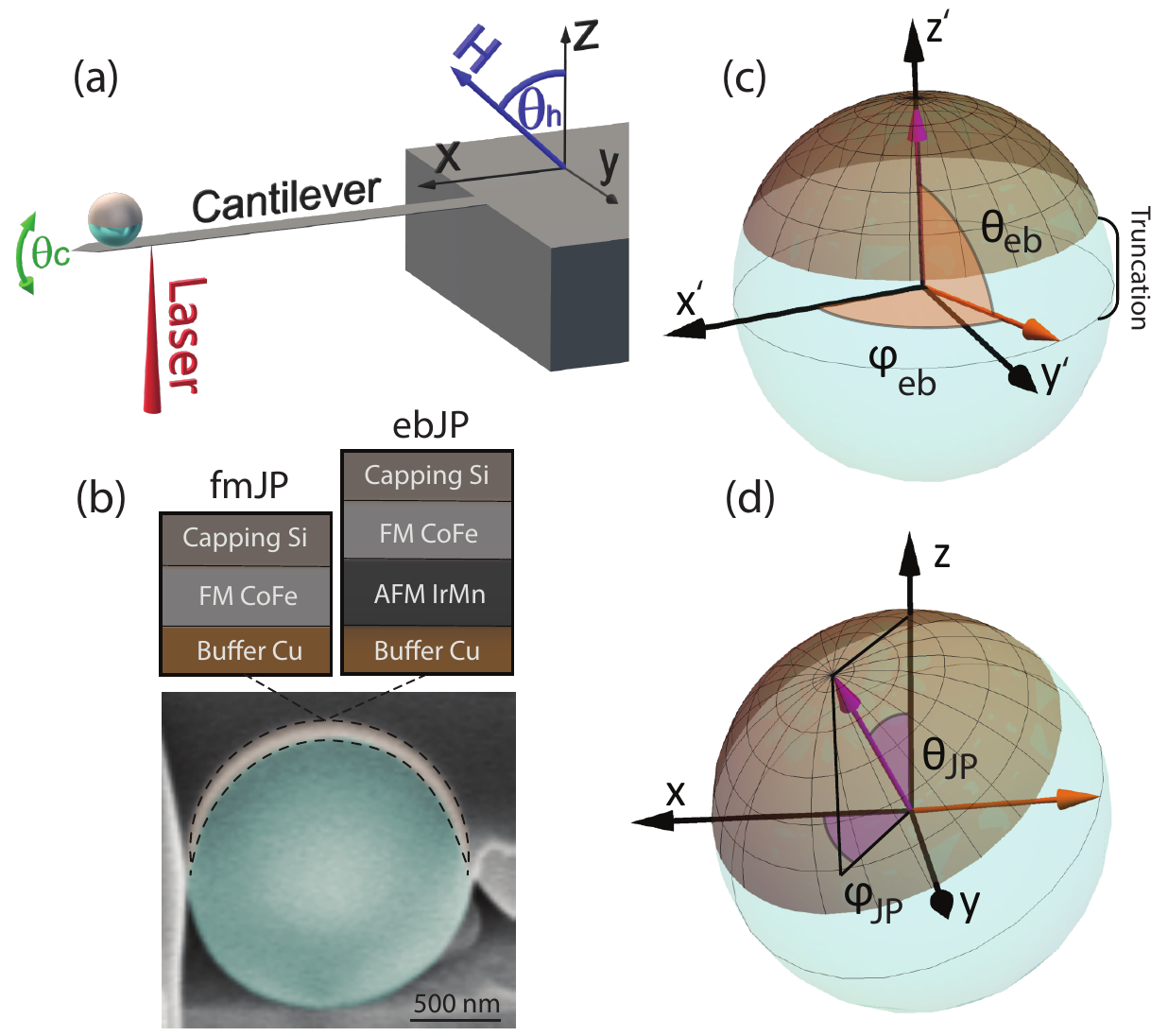}
  \caption{(a) Sketch of a cantilever with a JP attached to its tip
    and definition of the coordinate system.  (b) Cross-sectional SEM
    of a JP showing the gradient of the layer thickness. The two investigated layer stacks of the hemispherical cap are shown in the insets.  (c), (d)
    Definition of the angles setting the orientation of the
    unidirectional anisotropy vector used to mimic exchange bias
    effects ($\theta_{\text{eb}}$, $\varphi_{\text{eb}}$), and the
    angles defining the orientation of a JP on the cantilever
    ($\theta_{\text{JP}}$ and $\varphi_{\text{JP}}$).}
	\label{fig:setup}
\end{figure}
\begin{figure}[t]
	\centering
	\includegraphics[width=.48\textwidth]{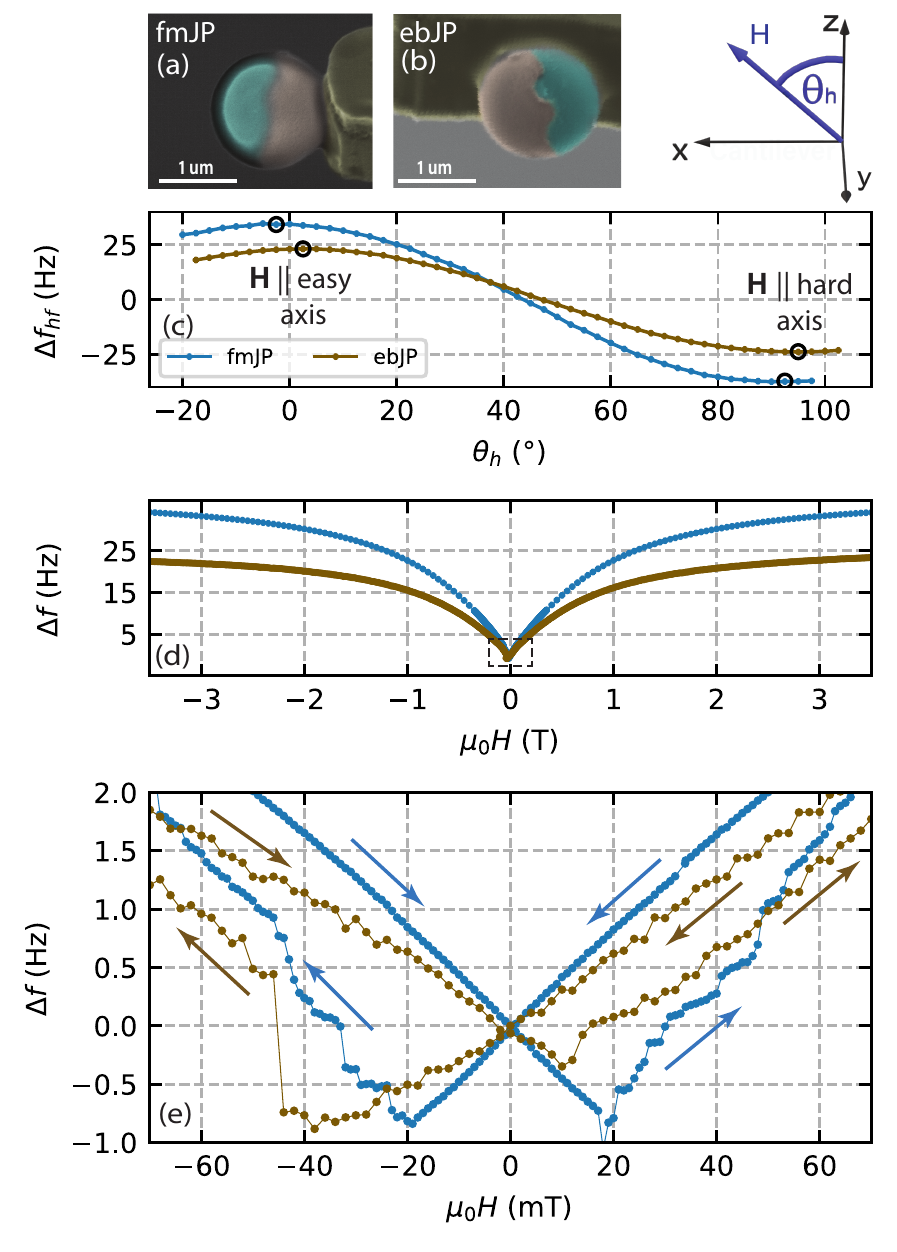}
	\caption{False color SEMs of the (a) fmJP and (b) ebJP
          attached to the tip of a cantilever, respectively. The
          coordinate system is shown on the right.  (c)
          $\Delta f_{\text{hf}}(\theta_h)$ measured at
          $\mu_0 H=\SI{3.5}{\tesla}$ for the fmJP (blue) and ebJP
          (brown).  Black circles indicate $\theta_h$ of the
          hysteresis measurements, which are shown in (d), (e), and Fig.~\ref{fig:simu}.}
	\label{fig:expdata}
\end{figure}

We fabricate the magnetic JPs by coating a self-assembled template of
$\SI{1.5}{\micro\meter}$-sized silica spheres with thin layers of
different materials via sputter-deposition.  The non-magnetic silica
spheres are arranged on a silica substrate using entropy
minimization~\cite{micheletto_langmuir_1995}, which allows the
formation of hexagonal close-packed monolayers.  JPs with two
different layer stacks, shown in Fig.~\ref{fig:setup} (b), are
produced.  Ferromagnetic JPs (fmJPs) are fabricated by depositing a
$\SI{10}{\nano\meter}$-thick Cu buffer layer directly on the silica
spheres, followed by a $\SI{10}{\nano\meter}$-thick layer of
ferromagnetic CoFe.  The film is sealed by a final
$\SI{10}{\nano\meter}$-thick layer of Si.  A second type of JP, which
we denote exchange-bias JPs (ebJPs), includes an additional
$\SI{30}{\nano\meter}$-thick afm layer of Ir$_{17}$Mn$_{83}$ between
the Cu buffer and the fm layer.  Layer deposition is performed by
sputtering in an external magnetic field of
$\SI{28}{\kilo\ampere/\meter}$ applied in the substrate plane, i.e.\
in the equatorial plane of the JPs, in order to initialize the
exchange bias by field growth. This fabrication process is described
in detail in Tomita et al.~\cite{tomita_magnetic_2021}.  Individual
JPs are then attached to the apex of a cantilever for magnetic
characterization in a last fabrication step, as shown in the scanning
electron micrographs (SEMs)
of Fig.~\ref{fig:expdata}~(a) and (b).\\

Note that the values given for thicknesses are nominal and that the
film thickness gradually reduces towards the equator of the sphere
with respect to the top, as shown in Fig.~\ref{fig:setup} (b), because
of the deposition process~\cite{tomita_magnetic_2021}.  Furthermore,
the touching points of the next neighbors in the hexagonal closed
packed arrangement of the silica spheres on the substrate template impose a
lateral irregularity on the equatorial line of the capping layers.
This is best seen in Fig.~\ref{fig:expdata} (a) and (b).\\

%
%
We measure the magnetic hysteresis of each an individual fmJP and an
individual ebJP via DCM.  DCM is a technique to investigate
individual, nano- to micrometer-sized magnetic specimens, similar to a
standard vibrating superconducting quantum interference device (SQUID)
magnetometer.  The key differences are that DCM is sensitive enough to
measure much smaller magnetic volumes than a vibrating SQUID
magnetometer and that it measures magnetic properties with respect to
rotations of the external magnetic field, rather than modulations of
its amplitude as in measurements of magnetic susceptibility.  Details
on the technique and measurement setup can be found in
Refs.~\onlinecite{gross_dynamic_2016,
  gross_meso_2021}.\\

\begin{figure}[t]
	\includegraphics[width=.48\textwidth]{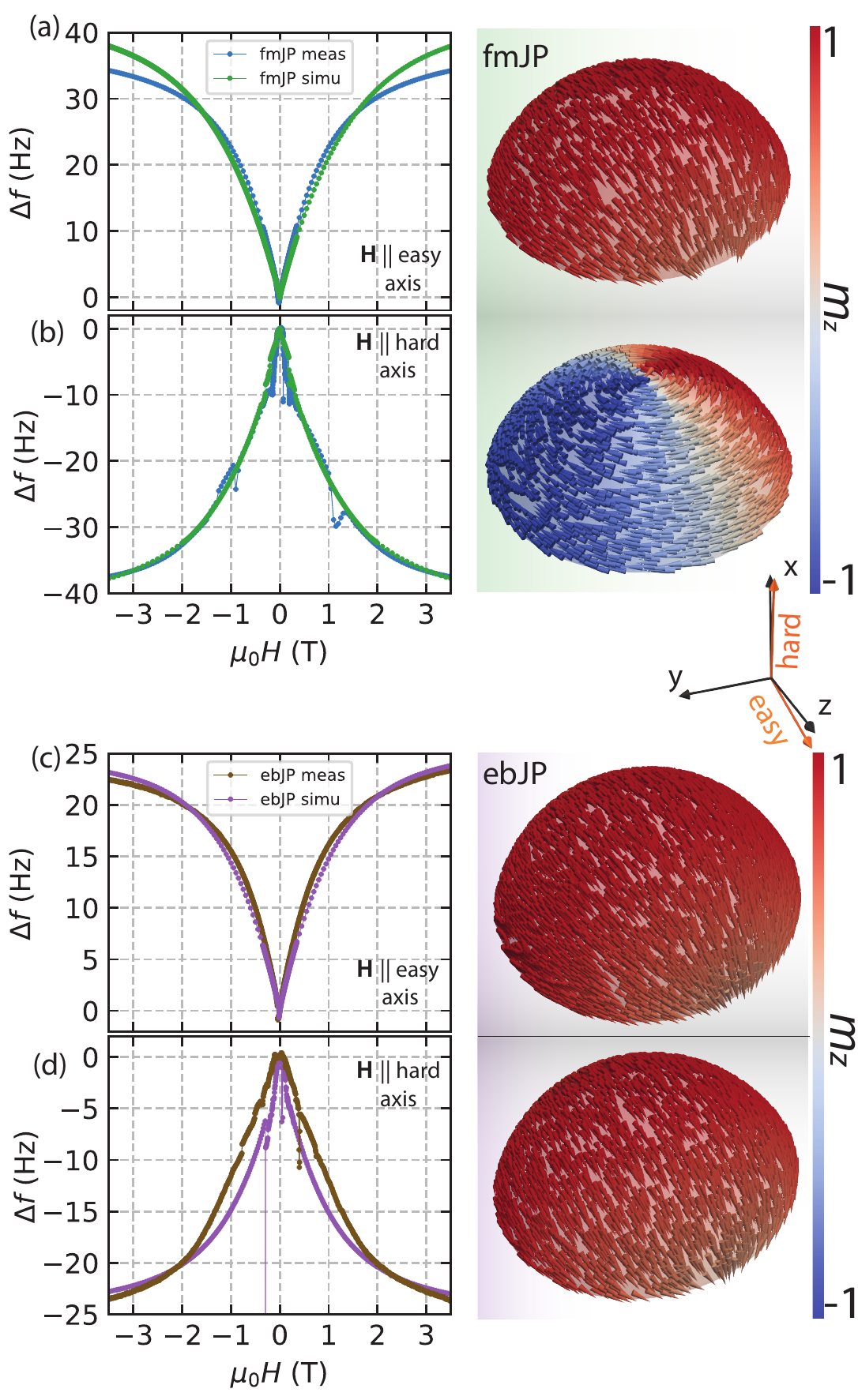}
	\caption{Measured and simulated $\Delta f(H)$ of the fmJP for $\mathbf{H}$ applied along the (a) easy and (b) hard axis, respectively. A visualization of each corresponding simulated remanent magnetic state is shown on the right. The same set of data for the ebJP is shown in (c) and (d).}
	\label{fig:simu}
\end{figure}

A magnetic specimen is attached to the tip of a cantilever, which is
driven in a feedback loop at its resonance frequency $f$ with a fixed
amplitude, actuated by a piezoelectric transducer.  A uniform external
magnetic field $\mathbf{H}$ is applied to set the magnetic state of
the specimen under investigation.  $\mathbf{H}$ can be rotated within
the plane perpendicular to the cantilever's rotation axis ($xz$-plane)
with a span of $\SI{117.5}{\degree}$ and a maximum field amplitude of
$H=\pm\SI{3.5}{\tesla}$.  Its orientation is set by the angle
$\theta_h$ as defined and indicated in Fig.~\ref{fig:setup} (a).  The
magnetic torque acting on the sample results in a deflection of the
cantilever as well as a shift in its resonance frequency, which is
given by
\begin{equation}
\label{eq:Df}
\Delta f = f-f_{0}=\frac{f_{0}}{2k_{0}l_{e}^{2}} \left( \frac{\partial^{2} E_{m}}{ \partial {\theta_{c}^{2}}}\bigg \vert_{\theta_{c}=0} \right).
\end{equation}
$f_{0}$ and $k_{0}$ are the resonance frequency and spring constant of
the cantilever at zero applied field, respectively, $l_{e}$ is the
effective length of the cantilever, $\theta_{c}$ the oscillation
angle, and $E_{m}$ the magnetic energy of the specimen. Properties of
the cantilevers used in the experiments can be found in section I of
the appendix.\\
In the limit of large applied magnetic field, such that the Zeeman
energy dominates over the anisotropy energy, all magnetic moments
align along $\mathbf{H}$.  In this limit $\Delta f$ asymptotically
approaches a value determined by the involved anisotropies, their
respective directions, the total magnetic moment, $\theta_h$, and the
mechanical properties of the
cantilever~\cite{gross_dynamic_2016,gross_meso_2021}.  In particular,
the determination of these asymptotes allows us to extract the
direction of magnetic easy and hard axes by measuring the angular dependence
of the frequency shift at high field $\Delta f_{\text{hf}}$.  The maximum positive (minimum negative)
$\Delta f_{\text{hf}}(\theta_h)$ indicates the easy (hard) axis.  For
these measurements, shown in Fig.~\ref{fig:expdata}~(c), we apply
$\mu_0 H=\SI{3.5}{\tesla}$, where we expect to be in the high field
limit.  See section VI and VII of the appendix for more details.  We
also measure the magnetic hysteresis of $\Delta f (H)$ by sweeping the
applied field $H$ up and down for $\theta_h$ fixed to the values determined for the magnetic easy and hard axes, respectively.  This procedure
reveals signatures of the JPs' magnetic reversal, as shown in
Figs.~\ref{fig:expdata}~(d),~(e) and \ref{fig:simu}.\\

In order to analyze both these types of measurements, we perform
micromagnetic simulations using the finite-element software
Nmag~\cite{fischbacher_nmag_2007}.  We calculate $\Delta f(H)$ from
the micromagnetic state for parameters set by the experiment
\cite{gross_dynamic_2016, mehlin_vortex_2018}.  Matching these
simulations to the measurements gives us a detailed understanding of
the progression of the magnetic configurations present in the JPs
throughout the reversal process.  Events, such as the nucleation of a
magnetic vortex, can be identified and associated with features in
$\Delta f (H)$ measured via DCM.\\

%
Fig.~\ref{fig:expdata} (a) and (b) shows false color SEMs of the
measured fmJP and ebJP, respectively, each attached to the tip of a
cantilever.
The orientation of the particles in the images can be correlated with
the angle $\theta_h$ of the maxima and minima found in the high-field
frequency shift $\Delta f_{\text{hf}}(\theta_h)$ in dependence of the
magnetic field angle $\theta_h$, shown in Fig.~\ref{fig:expdata} (c).
Doing so, we find a magnetic easy direction in the equatorial plane of
the particles and a hard direction along the axis of the pole.  The
$\SI{90}{\degree}$ angle between easy and hard direction is a clear
indication that uniaxial anisotropy is the dominant anisotropy in the
system.  We ascribe the latter to the shape of the JPs, because no
other strong anisotropies are expected.\\
The field-dependent frequency shift $\Delta f(H)$ for $\mathbf{H}$
aligned along the easy axis, see Fig.~\ref{fig:expdata}~(d), shows a
typical hysteretic, V-shaped curve, that approaches a horizontal
asymptote for high field magnitudes~\cite{gross_dynamic_2016}.  The
fmJP shows a symmetric asymptotic behavior for $\mu_0 H = \SI{3.5}{T}$
and $-\SI{3.5}{T}$ (blue curve).  Magnetic reversal at low fields,
$\mu_0 H$ around $\pm \SI{20}{\milli\tesla}$, is symmetric upon
reversal of the field sweep direction, as shown in
Fig.~\ref{fig:expdata}~(e).  This behavior is expected for a ferromagnetic
particle with a magnetic field applied along its easy axis.\\
In contrast, measurements of the ebJP reveal asymmetric asymptotic
behavior with $\Delta f_{\text{hf}}$ values differing by about
$\SI{0.9}{\hertz}$ for $\mu_0H=\pm \SI{3.5}{T}$, as seen in the brown
curve of Fig.~\ref{fig:expdata}~(d).  Furthermore, after a full
hysteresis cycle, we observe a reduction in the difference of
$\Delta f_{\text{hf}}$ at $\pm \SI{3.5}{\tesla}$ by about
$\SI{0.4}{\hertz}$, which is evidence for magnetic
training~\cite{buchter_magnetization_2015,Paul_2012_SymmetricMagnetizationReversal}.
Measurements of the ebJP also show a highly asymmetric magnetic
reversal, which occurs at $\mu_0 H = -\SI{44}{\milli\tesla}$ when
sweeping the field down and at $\mu_0 H = \SI{12}{\milli\tesla}$ when
sweeping the field up.  All of these findings are characteristic of an
exchange bias imposed on the
fm layer by the afm layer.\\

In order to draw conclusions about the magnetic state of the JPs, we
establish a micromagnetic model for each of the two types of JPs over
many iterations of comparison to measured $\Delta f (H)$ and
variations of the parameters for $\mathbf{H}$ applied along the
magnetic easy and hard axis, respectively.  This iterative process,
information from literature, and observations from SEMs lead to a
final set of parameters (see appendix, section I) and a model that
reproduces the measured $\Delta f (H)$.  The model assumes that the
magnetic JPs are made from a hemispherical shell with a thickness
gradient from the pole towards the equator, which accounts for the
gradual reduction of the shell thickness away from the pole, as shown
in Fig.~\ref{fig:setup}~(b).  The hemisphere is also
truncated~\cite{mourkas_curvature_2021} by a latitudinal belt around
the equator, reflecting observations from the SEM images in
Figs.~\ref{fig:expdata}~(a) and (b).  For simplicity, in the
simulations, we do not account for the magnetic film's irregular edge
at the equator and a possible change in the crystallographic texturing
with respect to the particle surface as a function of position within
the cap.  The orientation of a JP with respect to the cantilever
rotation axis and $\mathbf{H}$ is set by infering the orientation from
the SEMs and followed by an iterative tuning of the angles
$(\theta_{\text{JP}}, \varphi_{\text{JP}})$, as defined in
Fig.~\ref{fig:setup}~(d), to match the measured $\Delta f(H)$.\\
Figs.~\ref{fig:simu}~(a) and (b) shows agreement between the measured
and simulated $\Delta f(H)$ curves for the fmJP for $\mathbf{H}$
applied along the magnetic easy and hard axis, repectively.  Similar
curves are shown for the ebJP in Figs.~\ref{fig:simu}~(c) and (d).
Details on the progression of the magnetization configurations as a
function of $H$, as indicated by the simulations, are found in section
II of the appendix.  Here, we focus on the remanent magnetization
configurations in both types of particles, as shown on the right of
Fig.~\ref{fig:simu}.\\

For the fmJP, an onion state is realized after sweeping $\mathbf{H}$
along the easy axis while a global vortex state is found after
sweeping along the hard axis.  In applications, magnetic JPs are
subject to considerable disturbances from the outside, including
thermal activation, interactions with other nearby magnetic particles,
and alternating external magnetic fields for actuation.  Hence, we can
expect a magnetic JP to relax to its ground state configuration over
time.  The simulation of the fmJP shows that when the magnetization is
in the global vortex state, its magnetic energy is
\SI{101}{\atto\joule} lower than when it is in the onion state.  The
global vortex state is therefore energetically more favorable than the
onion state, which is also true if compared to any other remanent
state that we have found in simulations for fmJPs, as discussed in
appendix, section IV.  This analysis suggests that a remanent global
vortex state, which has a vanishing total magnetic moment, is realized
in fmJPs over time, independent of magnetic history.  If we normalize
the magnetic moment of this state by the saturation moment, $M_sV$, we
find that the global vortex state hosted by the fmJP has a moment
value of 0.03, precluding the use of such particles in applications.\\

We establish a similar micromagnetic model for the ebJP.  To model the
effect of an exchange bias imposing a preferred direction on the
magnetic moments in the ferromagnet, a unidirectional anisotropy is
added to the simulation. 
Note that other influences of the afm layer are not accounted for, especially contributions to the coercive
field~\cite{Mueglich_2016_TimeDependent}, rotational anisotropies~\cite{Geshey_2002_Rotatable}, or contributions arising due to its granular structure \cite{merkel_interrelation_2020}. 
For this reason, neither the asymmetric values of the magnetic
reversal fields nor the observed training effects are correctly
reproduced by the simulation.  The unidirectional anisotropy, described by a unit vector $\mathbf{\hat{u}}_{eb}$ with orientation angles $(\theta_{eb},\varphi_{eb})$, as shown in Fig.~\ref{fig:setup}~(c), and an anisotropy constant $K_{eb}$, is expected to lie somewhere in the equatorial plane of the ebJP. The specific orientation of $\mathbf{\hat{u}}_{eb}$ within this plane is induced in the afm by the magnetic field applied during deposition. Note that for simplicity, $K_{eb}$ is kept constant within the whole volume of the magnetic cap, despite thickness variations of the afm layer with $\theta_{eb}$.\\

The knowledge of where $\mathbf{\hat{u}}_{eb}$ points within the
equatorial plane is lost after attaching the JP to the cantilever.  In
the simulations, we choose to align $\mathbf{\hat{u}}_{eb}$ along the
direction in the equatorial plane that coincides with the applied
external field in the easy-axis configuration, even though it could
point along any direction in this plane.  This assumption that
$\mathbf{H}$ and $\mathbf{\hat{u}}_{eb}$ are collinear in the
easy-axis measurements means that our simulations predict the maximum
possible $\Delta f$ for any given choice of the unidirectional
anisotropy constant $K_{eb}$ and hence give a lower bound for
$K_{eb}$.

We adjust this anisotropy to match the measured
$\Delta f(H)$ along both the easy and hard axes and find good
agreement for $K_{eb}=\SI{22.5}{\kilo\joule/\meter\cubic}$.  In
particular, the model reproduces the asymmetry in
$\Delta f_{\text{hf}}$ for both the easy and hard-axis alignments, as
shown in Fig.~\ref{fig:simu}~(c) and (d).  This value of $K_{eb}$ is
also consistent with results from
Ref.~\onlinecite{merkel_interrelation_2020}, once the influence of a
reduced sample size is considered~\cite{nogues_exchange_2005}.\\

At remanence, the simulations show that the ebJP hosts an onion state
irrespective of its magnetic history, as shown in
Figs.~\ref{fig:simu}~(c) and (d).  To exclude the presence of an
equilibrium global vortex state, we test this state's stability by
initializing the ebJP in a global vortex state at remanence and then
relaxing the system to a local energetic minimum.  Following this
procedure, the system relaxes to the onion state.  As a result, we can
exclude the global vortex state as a possible equilibrium remanent
state in this ebJP.\\

For the simulations of the ebJP we find a total magnetic moment at
remanence, normalized by its maximum value of $M_sV$, of 0.89 and 0.71
depending on whether $\mathbf{H}$ is applied along the hard or easy
direction of the external field, respectively.  This remanent moment
represents an increase of more than one order of magnitude compared to
the remanent moment of the fmJP.  Hence, introducing exchange bias to
magnetic JPs, if strong enough, succeeds in stabilizing a high-moment
onion state in remanence.\\

%
%
To conclude, micrometer-sized JPs capped with an antiferromagnetic/ferromagnetic or
purely ferromagnetic thin film system have been mass produced through a
sputter-deposition process.  We have investigated the magnetic
reversal and remanent magnetic configurations of individual specimens of these JPs
using DCM and corresponding micromagnetic simulations.  Although the
fmJPs host a global vortex state in remanence with a vanishing
magnetic moment, the addition of an antiferromagnetic layer in ebJPs
successfully changes the remanent configuration to a stable
high-moment onion state.  Unlike previous measurements on close packed particle arrays, our
measurements on individual JPs show that the stability of this
high magnetic moment texture in remanence is a property of the individual
particles and present in absence of interparticle interactions.\\

%
We thank Sascha Martin and his team in the machine shop of the Physics
Department at the University of Basel for help building the
measurement system.  We acknowledge the support of the Canton Aargau
and the Swiss National Science Foundation under Grant
No. 200020-159893, via the Sinergia Grant Nanoskyrmionics (Grant
No. CRSII5-171003), and via the National Centre for Competence in
Research Quantum Science and Technology. Calculations were performed
at sciCORE (http://scicore.unibas.ch/) scientific computing core
facility at University of Basel.\\

The data that support the findings of this study are available from the corresponding author upon reasonable request.

\bibliographystyle{unsrt}
\end{document}


\title{Supplementary Information: Magnetic hysteresis of individual Janus particles with hemispherical exchange biased caps}
\author{S.~Philipp} \affiliation{Department of Physics, University of Basel, 4056 Basel, Switzerland}
\author{B.~Gross} \affiliation{Department of Physics, University of Basel, 4056 Basel, Switzerland}
\author{M.~Reginka}\affiliation{Institute of Physics, University of Kassel, 34132 Kassel, Germany}
\author{M.~Merkel}\affiliation{Institute of Physics, University of Kassel, 34132 Kassel, Germany}
\author{M.~Claus} \affiliation{Department of Physics, University of Basel, 4056 Basel, Switzerland}
\author{M.~Sulliger} \affiliation{Department of Physics, University of Basel, 4056 Basel, Switzerland}
\author{A.~Ehresmann}\affiliation{Institute of Physics, University of Kassel, 34132 Kassel, Germany}
\author{M.~Poggio} \affiliation{Department of Physics, University of  Basel, 4056 Basel, Switzerland} \email{martino.poggio@unibas.ch}
\homepage{http://poggiolab.unibas.ch/}
\maketitle

\newpage
\section{Cantilever properties and simulation details}
\label{sec:canti}
Cantilevers are fabricated from undoped Si. They are
\SI{75}{\micro\meter}-long, \SI{3.5}{\micro\meter}-wide,
\SI{0.1}{\micro\meter}-thick with a mass-loaded end and have a
\SI{11}{\micro\meter}-wide paddle for optical position detection. The
resonance frequency $f_0$ of the fundamental mechanical mode used for
magnetometry is on the scale of a few kilohertz.  Spring constant
$k_0$ and effective length $l_e$ are determined using a finite element
approximation~\cite{comsol}.
For the cantilever used for the fmJP we find $f_0=\SI{5285.8}{\hertz}$, $k_0=\SI{249}{\micro\newton/\meter}$ and $l_e=\SI{75.9}{\micro\meter}$. For the cantilever of the ebJP $f_0=\SI{5739.3}{\hertz}$, $k_0=\SI{240}{\micro\newton/\meter}$ and $l_e=\SI{75.9}{\micro\meter}$.\\

Micromagnetic simulations are performed with the finite-element
software package Nmag~\cite{fischbacher_nmag_2007}.  As an
approximated geometry for the JPs a semi-sphere shell with a thickness
gradient from the pole towards the equator is used, that is truncated
at the equator, as discussed in the main text. The exchange constant
is set to
$A_{ex} =
\SI{30}{\pico\joule/\meter}$~\cite{berkov_micromagnetic_2011}.

In case of the fmJP, opposing to the SEM image in Fig.~2 (a) in the
main text, which suggests a truncation of the fm layer by about
\SI{250}{\nano\meter}, it needs to be set to \SI{350}{\nano\meter} or
even more to match the high field progression of $\Delta f(H)$. For
the same reason the nominal thickness of \SI{10}{\nano\meter} of the
fm layer needs to be increased to at least \SI{12}{\nano\meter} at the
pole, which is then gradually reduced to 0 at the equator. These two
geometric constraints are necessary to keep $M_s$ at a reasonable
value below the bulk value of
\SI{1.95}{\mega\ampere/\meter}~\cite{coey_magnetism_2010}. This
suggests, that significantly more fm material than anticipated is
deposited on the region around the pole of the JPs, which is the most
directly exposed area of the sphere during deposition.

For the simulation of the fmJP we set the following parameters:
saturation magnetization $M_s=\SI{1.8}{\mega\ampere/\meter}$, silica
sphere diameter of \SI{1.5}{\micro\meter}, truncation
$d=\SI{350}{\nano\meter}$, particle orientation
$(\theta_{JP},
\varphi_{\text{JP}})=(\SI{91}{\degree},\SI{2}{\degree})$ and maximum
allowed mesh cell size \SI{7.5}{\nano\meter}.

For the ebJP, the geometric parameters have to be adjusted less from
their nominal values than for the fmJP, in order to match between the
micromagnetic model to the experiment. This result suggests that the
afm layer, which is deposited before the fm, acts as an adhesive for
the fm, and the ebJP is coated more homogeneously than the fmJP.

For the simulation of the ebJP we set the following parameters:
$M_s=\SI{1.44}{\mega\ampere/\meter}$, fm layer thickness of
\SI{10}{\nano\meter} at the pole, gradually reduced to 0 at the
equator, silica sphere diameter of \SI{1.5}{\micro\meter},
$d=\SI{350}{\nano\meter}$,
$(\theta_{JP},
\varphi_{\text{JP}})=(\SI{85}{\degree},\SI{10}{\degree})$, maximum
allowed mesh cell size \SI{7.0}{\nano\meter},
$(\theta_{\text{eb}},
\varphi_{\text{eb}})=(\SI{-90}{\degree},\SI{0}{\degree})$,
unidirectional anisotropy constant
$K_{eb}=\SI{22.5}{\kilo\joule/\meter\cubic}$.

For the generic simulations in sections~\ref{sec:generic} and
\ref{sec:shape} we have used $M_s=\SI{1.8}{\mega\ampere/\meter}$, , fm
layer thickness of \SI{10}{\nano\meter} at the pole, gradually reduced
to \SI{1}{\nano\meter} at the equator,a silica sphere diameter of
\SI{500}{\nano\meter},
$(\theta_{JP},
\varphi_{\text{JP}})=(\SI{0}{\degree},\SI{0}{\degree})$, and
$(\theta_{\text{eb}},
\varphi_{\text{eb}})=(\SI{-90}{\degree},\SI{0}{\degree})$. Parameters
that are not mentioned here are given in the main text.
\section{Progression of the magnetic state with external field}
\label{sec:progression}
\subsection{Ferromagnetic Janus particles}
\label{sec:fmjp}
\begin{figure*}[t]
	\centering
	\includegraphics[width=15.5cm]{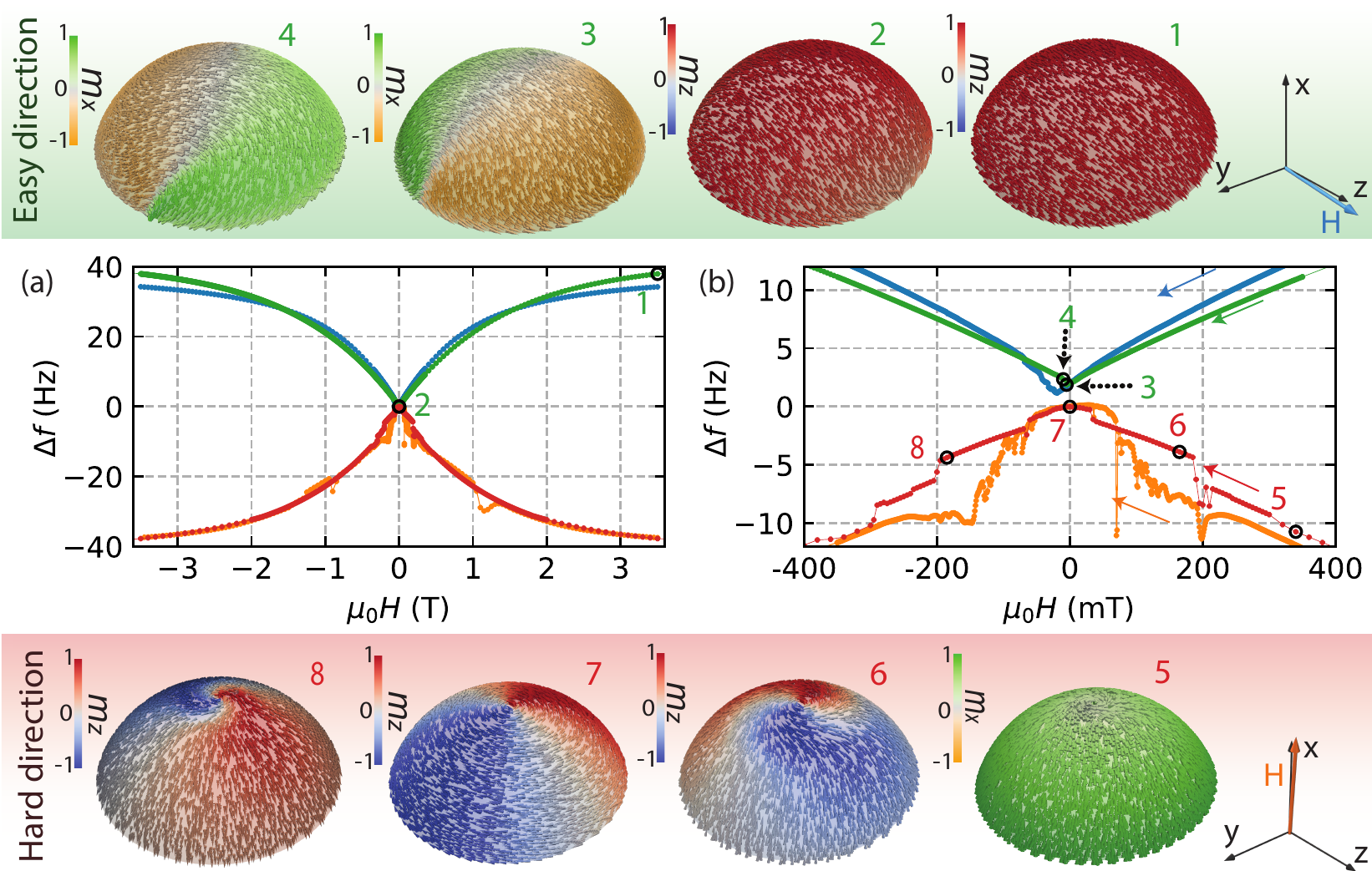}
	\caption{Data for the ferromagnetic Janus particle. (a)
          Measured $\Delta f(H)$ for easy (blue) and hard (orange)
          alignment of the external field, as well as the simulated
          magnetic configurations in green (easy) and red (hard). (b)
          Close-up of (a) for low fields. Data with easy orientation
          is offset by \SI{2}{\hertz} for better visibility. Colored
          arrows indicate field sweep directions. Numbers in (a) and
          (b) denote the field values for the configurations of the
          magnetic state.}
	\label{fig:fmjp}
\end{figure*}

$\Delta f(H)$, measured for $\mathbf{H}$ parallel to the magnetic easy
(blue data) and hard axis (orange data), respectively, as shown in
Fig.~\ref{fig:fmjp}~(a), gives direct information on high field
behavior and magnetic reversal of the JPs.
For $\mathbf{H}$ parallel to the magnetic easy axis, an overall V-shape suggest Stoner-Wohlfarth like behavior for most of the field range in Fig.~\ref{fig:fmjp}~(a). As seen in the close-up in Fig.~\ref{fig:fmjp}~(b), magnetic reversal appears to take place through a few sequential switching events at small negative reverse fields.\\
The simulated $\Delta f(H)$, also shown in Fig.~\ref{fig:fmjp} (green
points) together with a few exemplary configurations of the simulated
magnetic state of the JP, can give more insight into what happens
during the field sweep. Starting from full saturation, most magnetic
moments stay aligned with the easy direction down to very low reverse
fields, nicely seen in Fig.~\ref{fig:fmjp}, configuration 1 at
\SI{3.5}{\tesla} and configuration 2 at remanence. The latter is an
onion state. This progression of configurations is consistent with the
Stoner-Wohlfarth-like behavior of the experimental $\Delta f(H)$.
Magnetic reversal takes place through the occurrence of a so-called
S-state, for which the magnetization follows the curvature of an S. The reversal is shown in configurations 3 and 4 in Fig.~\ref{fig:fmjp}. 
Then, until full saturation is reached in reverse field, only
magnetic moments in proximity to the equator of the JP are slightly
canted away from the direction of the external field (and the easy
plane). This progression is robust in simulation, even though
sometimes, depending on slight variations of simulation parameters, a
vortex appears in reverse field instead of the S-state.
The observation of several, individual switching events during magnetic reversal in experiment may originate in vortex hopping, or switching of different regions in the JP due to variations in material and geometric parameters. Magnetic reversal through a vortex rather than an S-state may also explain the big difference of the coercive fields between experiment ($H_c\approx\SI{32}{\milli\tesla}$) and simulation ($H_c=\SI{6}{\milli\tesla}$).\\

For $\mathbf{H}$ parallel to the magnetic hard axis, the experimental data has an inverted V-shape, and there is no easily identifiable sign of magnetic reversal. Yet, for relatively large fields, around \SI{1.5}{\tesla}, switching events are observed, and exist up to negative fields of similar magnitude.
Typically, W-shaped curves are observed for measurements with the field aligned with the hard direction, and can be understood in a simple Stoner-Wohlfarth model, which is discussed in section \ref{sec:SWmodel}. The inverted V-shape instead of the W-shape is a peculiarity of the fact that the fm layer of the JP is curved everywhere. The angle between the local surface normal and the direction of the external magnetic field is different for every polar coordinate of the JP, which leads to a dependence of the local demagnetizing field on the polar coordinate. In consequence, the magnitude of the external magnetic field, for which the local magnetic moments start to rotate towards their local easy direction depends strongly on the position in the magnetic cap. This leads to the observed curve shape of $\Delta f(H)$, a more detailed discussion can be found in section~\ref{sec:shape}.
The magnetic progression in simulation for external field alignment
with the hard direction can be summarized as follows: Starting from
full saturation, the magnetic moments start rotating towards the easy
plane with decreasing field magnitude due to the competition between
shape anisotropy and Zeeman energy. This takes place for different
magnitudes of $\mathbf{H}$ depending on where a magnetic moment is
located in the JP, as discussed earlier. Configuration 5 in
Fig.~\ref{fig:fmjp} shows a state for which magnetic moments at the
pole have already started to rotate, while magnetic moments in
proximity to the equator remain aligned with the external field.
Superimposed to this rotation, a minimization of the system's energy
by formation of a magnetic vortex localized at the pole for
around \SI{1.8}{\tesla} takes place, which grows in size with decreasing field,
see configuration 6. In simulation, this is a gradual evolution, and only for fields below
about \SI{300}{\milli\tesla} jumps in $\Delta f$ due to vortex
movement are observed. This process is in contrast to the
discontinuities that occur at around \SI{1.5}{\tesla} in the
experiment, but can be explained by vortex hopping from pinning site
to pinning site.  The latter may be present due to fabrication
inhomogeneities in the JPs~\cite{rossi_mfm_2019}. For zero field the
vortex dominates the magnetic configuration of the JP and has evolved
into a global vortex state, as shown in configuration 7. Configuration
8 shows the vortex in reverse field, which has changed polarity, and
has jumped to a slightly off-centered position. The latter is too
small to be visible in the figure.  Further decreasing the field, the
vortex sits centrally in the JP and shrinks in size, and vanishes
around \SI{-1.82}{\tesla}. At the same time, the magnetic moments
rotate towards the field direction depending on their position in the
JP, as described earlier.  Note, that by slightly changing simulation
parameters, we find that features due to vortex entrance and hopping
may manifest themselves in $\Delta f (H)$ with strongly differing
magnitude and for different field values. Introducing artificial
pinning cites in simulation can be used to adjust the vortex hopping
to match the observed signals more precisely~\cite{rossi_mfm_2019},
but consumes vast amounts of computational time and should still be
understood only as an exemplary progression of the magnetic state.
\subsection{Exchange-biased Janus particles}
\label{sec:ebjp}
\begin{figure*}[t]
	\centering
	\includegraphics[width=15.5cm]{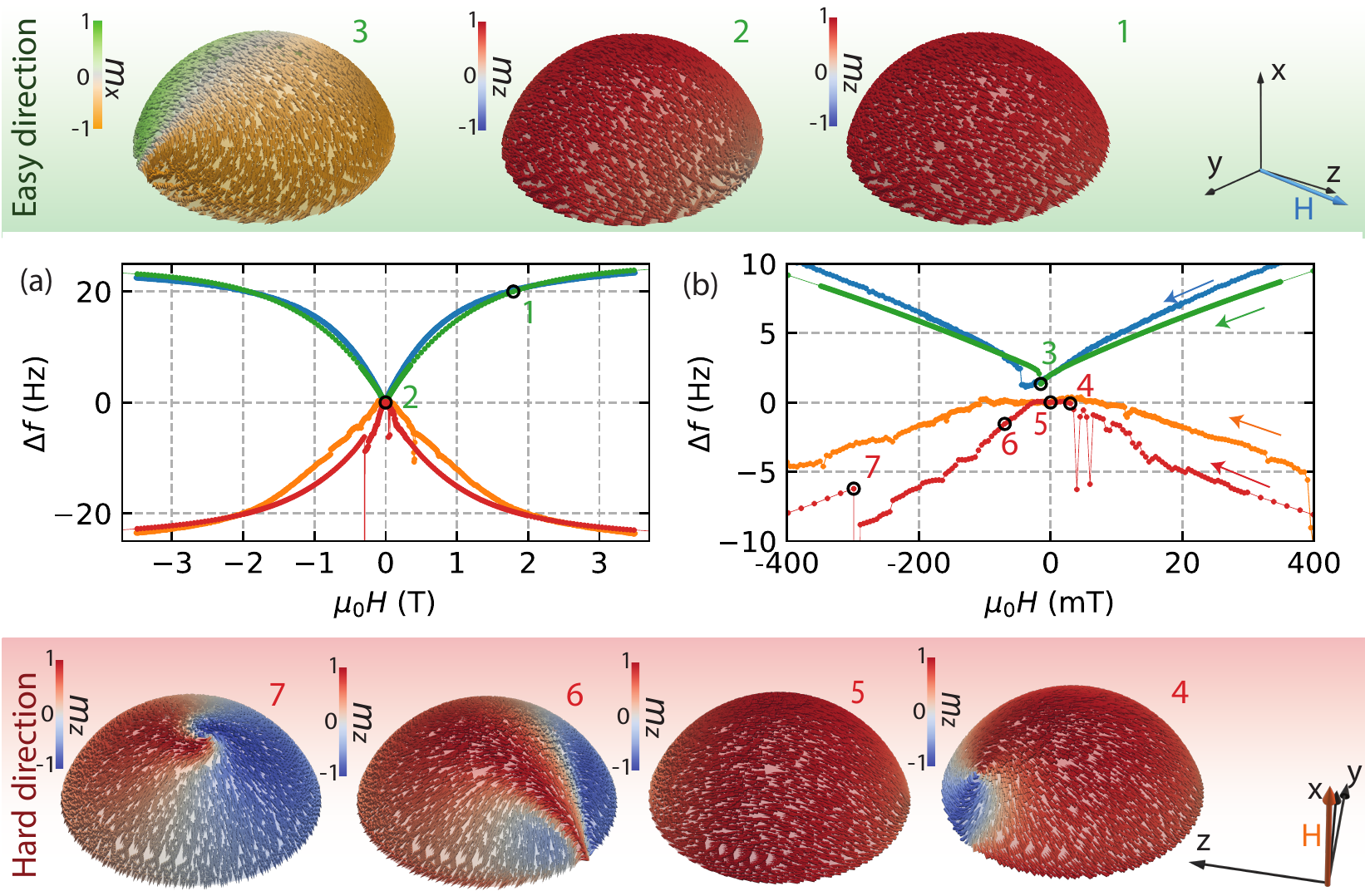}
	\caption{Data for the exchange biased Janus particle. (a)
          Measured $\Delta f(H)$ for easy (blue) and hard (orange)
          alignment of the external field, as well as the simulated
          pendants in green (easy) and red (hard). (b Close-up of (a)
          for low fields. Data with easy orientation is offset by
          \SI{2}{\hertz} for better visibility. Numbers in (a) and (b)
          indicate the field values for the configurations of the
          magnetic state shown in (c) for hard and in (d) for easy
          alignment.}
	\label{fig:ebjp}
\end{figure*}
The progression of the magnetic state for the ebJP is very similar to
the fmJP, yet, there are crucial differences. See Fig.~\ref{fig:ebjp}
for the DCM data, simulation results, and configurations of some
magnetic states.
%
%
For the field oriented in the magnetic easy direction the nearly polarized state, shown in Fig.~\ref{fig:ebjp} configuration 1, is similar to that shown in Fig.~\ref{fig:fmjp}, configuration 1. Reducing the field down to remanence, as shown in Fig.~\ref{fig:ebjp} configuration 2, we find an onion state just as for the fmJP. Magnetic reversal occurs again through an S-state, rather than via vortex formation, as shown in configuration 3. However, the reversal is shifted towards negative fields, and occurs for \SI{-15}{\milli\tesla} for the down sweep, and for \SI{-17.5}{\milli\tesla} for the up sweep of the magnetic field. This does not match the experimentally observed values, especially for the latter case, for which the switching occurs for positive field. This is no surprise, since the employed model does not account for the contribution of the exchange bias to the coercivity. Yet, both simulation and experiment show a shift of the hysteresis loop towards negative fields as compared to the fmJP.\\
We only observe a single switching event in experiment for the magnetic reversal, which is consistent with the behavior of the S-state in simulation. 
 For the alternative magnetic reversal process through vortex formation, we would expect several switching events due to vortex hopping. We find such a situation e.g. for a few reversal processes without unidirectional anisotropy, where geometrical parameters of the JPs have been varied, see section \ref{sec:generic}. Yet, it is also possible, that a strong pinning site favors the formation of a vortex, and keeps it in place for all field magnitudes up to the reversal point.\\
 If the magnetic field is swept from negative saturation up to remanence (not shown here), an onion state is present, that has its total magnetic moment pointing opposite to the exchange bias direction. For applications, this is an undesirable state. It is energetically less favorable than the state of parallel alignment, and if the energy barrier between the two states is overcome by an external influence, the JP will switch.\\

If the external field is aligned with the hard direction, and starts
at full saturation, the magnetic moments rotate towards the easy plane
depending on their position in the JP for decreasing field just as for
the fmJP. Further, the same, superimposed vortex formation takes
place, starting for $\SI{1.34}{\tesla}$. Again the vortex occupies
more and more volume of the JP with further decreasing field.

Superimposed, a local vortex forms at the pole of the ebJP for an
applied field of $\SI{1.34}{\tesla}$. As for the fmJP, the vortex
occupies more and more volume of the JP with further decreasing field.
However, upon further reducing the field, the vortex, rather than
inhabiting the whole JP as a global vortex centered at the pole of the
fmJP, it prefers to move to the side of the ebJP, as shown in
configuration 4 of Fig.~\ref{fig:ebjp}. Moving down from the pole
towards the equator, the vortex exits from the JP through the equator
for \SI{5}{\milli\tesla}, and an onion state is formed at remanence,
as shown in configuration 5. The orientation of the onion state is
governed by $\mathbf{\hat{u}}_{eb}$.  For a small reverse field a
domain wall state forms, as shown in configuration 6. With further
decreasing field, the domain wall is rotated with respect to the polar
axis of the JP. This state seems to be a precursor of the vortex
state, and the wall is subsequently replaced by the vortex, sitting
again in the center of the JP, as shown in configuration 7. The vortex
vanishes for -\SI{1.36}{\tesla}. Whether such a domain wall state is
indeed realized in the ebJPs for reverse fields, or if a vortex enters
from the equator and moves back to the center of the JP, as seen for
simulations of smaller JPs (see section \ref{sec:generic}), remains an
open question. The DCM signal shows in both experiment and simulation
many irregularities for the lower field range, which does not allow us
to draw clear conclusions on the magnetic state present in the
JPs. Nevertheless, the simulations clearly suggest that an onion state
should be realized at remanence, irrespective of the states present
during the hysteresis. This situation is markedly different than that
of the fmJP and is a direct consequence of the presence of exchange
bias.
\section{Stoner-Wohlfarth model for easy plane type anisotropy}
\label{sec:SWmodel}
The shape of the magnetic material of the fmJPs is, at least in a first approximation, rotationally symmetric around the pole axis. Further, the thickness gradient of the CoFe layer, as shown in the cross-sectional SEM in Fig.~1~(b) in the main manuscript, suggest that the magnetic material is concentrated in proximity to the pole, and that there is less material towards the equator. This material distribution suggests that a uniaxial anisotropy of easy plane type is imposed on the sample by its shape. 

\begin{figure*}[t]
  \includegraphics[width=17.2cm]{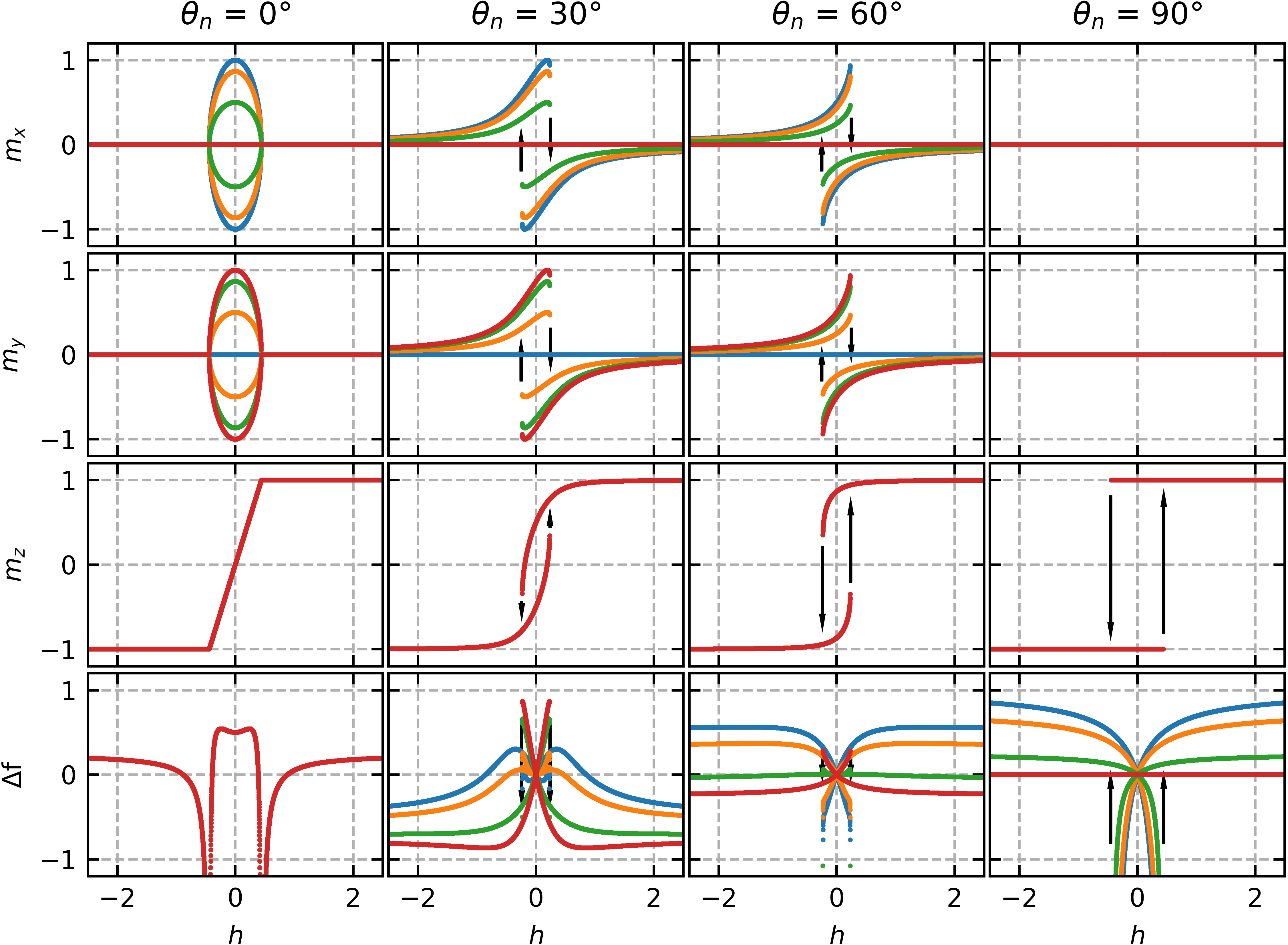}
  \caption{Components of $\mathbf{M}$ normalized by $M_s$ (first 3 rows) and $\Delta f$ normalized by $\frac{f_0 \mu_0 V M_s^2}{2 k_0 l_e^2}$ (last row) vs.\ normalized magnetic field $h=\frac{H}{M_s D_u}$ for different orientations of the anisotropy axis. Valid for uniaxial anisotropy with $D_u > 0$ (here $D_u=0.5)$. $\theta_u$ is increased from $\SI{0}{\degree}$ in the first column by $\SI{30}{\degree}$ per column up to $\SI{90}{\degree}$. $\phi_u$ is changed in the same steps, given by the different, color-coded graphs within each column (blue for $\phi_u=\SI{0}{\degree}$, orange for $\SI{30}{\degree}$, green for $\SI{60}{\degree}$ and red for $\SI{90}{\degree}$). Arrows indicate switching of the magnetization. $\Delta f$ has been scaled by a factor 0.5 and offset by 0.5 for $\theta_u=\SI{0}{\degree}$, and scaled by 2 for the other values of $\theta_u$.}
\label{fig:SWeasyplane}
\end{figure*}

The most basic approach to describe such a system is a Stoner-Wohlfarth model, in which a single macro magnetic moment replaces the ensemble of distributed magnetic moments \cite{morrish_physical_2001}. Following Ref.~\onlinecite{gross_dynamic_2016}, it is straight forward to calculate the magnetic hysteresis and connected DCM response for easy plane anisotropy, where the latter is manifested in a positive, effective demagnetizing factor $D_u$, opposing to a negative $D_u$ for easy axis anisotropy. The model is a good approximation for high fields, where all magnetic moments are aligned with the external field, and essentially behave like a single macro spin. At lower fields deviations from the curve of the SW-model indicate inhomogeneous spin orientation. We show the magnetic hysteresis of the components $m_x$, $m_y$ and $m_z$ of the normalized magnetization $\mathbf{m}$ and the DCM response $\Delta f$ in Fig.~\ref{fig:SWeasyplane} for several different orientations ($\theta_u$, $\phi_u$) of the uniaxial anisotropy axis $\mathbf{\hat{u}}$, which is perpendicular to the easy plane. The external field $\mathbf{H}$ has to be fixed in the $z$-direction in the model for technical reasons, which is why $\mathbf{\hat{u}}$ is varied rather than $\mathbf{H}$, contrary to the situation in experiment. Hence, the angles $\theta_{JP}-\theta_{h}$ and $\phi_{JP}$, describing the equivalent situation in experiment, have to be compared to $\theta_u$ and $\phi_u$, respectively. However, this does not limit the validity of the model.\\
For $\theta_u=\SI{0}{\degree}$, for which $\mathbf{H}$ is perpendicular to the easy plane, we find the typical W-shape for $\Delta f$ for a magnetic hard orientation of $\mathbf{H}$ \cite{gross_dynamic_2016}. The columnar arrangement of the components of the magnetization $m_x$, $m_y$ and $m_z$ together with $\Delta f$ allow to correlate changes in magnetic behavior with features in $\Delta f$, such as e.g. the transition from field alignment to the beginning of a rotation of $\mathbf{m}$ towards the easy axis.\\
For $\theta_u=\SI{90}{\degree}$, for which $\mathbf{H}$ lies in the easy plane, the typical V-shaped curve for an easy orientation of $\mathbf{H}$ is found, just as expected~\cite{gross_dynamic_2016}. This V-shape is connected to a perfect square hysteresis of $\mathbf{m}$. Intermediate values of $\theta_u$ lead to intermediate curve progression, which are a combination of the two extrema described above.

Compared to experiment (see Figs.~2 and 3 in the main manuscript) we find, that while in the case of the external field $\mathbf{H}$ in the easy plane (last panel in the last row in Fig.~\ref{fig:SWeasyplane}) the model generates relatively similar curve shapes for $\Delta f$, this is not true for $\mathbf{H}$ perpendicular to the easy plane (first panel in the last row). The vertical asymptotes for $h \approx \pm 0.5$ are missing in experiment. Further, the horizontal asymptote at high field is approached from negative rather than positive values, as it is seen in experiment. A more detailed analysis in the following section will show, that the latter originates in the curvature of the magnetic shell, which cannot be captured by a single spin model. Even though typically a good starting point, the SW model seems to be of rather limited validity for the description of the  relatively specific geometry of a spherical cap.
\section{Shape anisotropy of a truncated spherical halfshell}
\label{sec:shape}
\begin{figure}[t]
  \includegraphics[width=8cm]{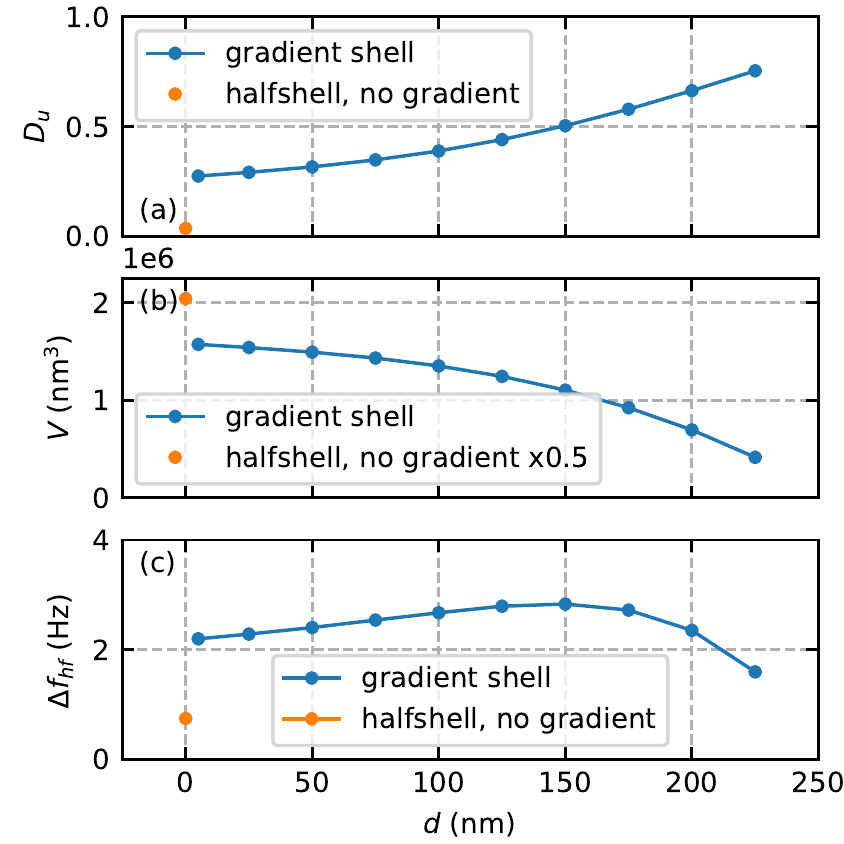}
  \caption{ (a) Effective demagnetization factor $D_u$ of a truncated spherical halfshell with a gradient shell thickness in dependence of the truncation $d$ (blue) and of a full halfshell (orange). (b) Volume and (c) high field frequency shift $\Delta f_{hf}$ of the same geometries as in (a).}
\label{fig:JP_generic_Du}
\end{figure}

As a measure for the strength of the shape anisotropy present in the JPs, and as a parameter used for SW-modeling, the knowledge of the effective demagnetization factor $D_u$ of the geometry is of value. If possible, $D_u$ is determined using an analytical expression~\cite{Hubert2014MagneticDT}, which, however, is not known for the given geometry. Using micromagnetic simulations as discussed in the main text, we can extract a good approximation to the demagnetization factor of a given geometry, without necessity for an analytical formula. Here, we analyze a generic, truncated spherical halfshell as defined in section \ref{sec:canti} for different degrees of truncation $d$.\\
We quickly recall the definition of the effective demagnetization factor $D_u = D_z - D_x$, where $D_z$ and $D_y=D_x$ are the demagnetization factors of a magnetic object that is rotationally symmetric in the $xy$ plane. $-0.5 < D_u < 1$ is true, and $D_u<0$ is valid for prolate and $D_u>0$ for oblate bodies. $D_u=0$ is the case for a perfectly spherical body. We find a minimum $D_u$ of approximately 0.25 for the smallest truncation, and $D_u$ increases with truncation as shown in Fig.~\ref{fig:JP_generic_Du} (a). Hence, shape anisotropy gets stronger as $d$ is increased, which can be roughly understood as the transformation from a spherical halfshell to a disc.
Note, that a spherical halfshell without thickness gradient and without truncation leads to a $D_u$ just slightly large than zero. This implies, that pointing in the easy plane is not very much more favorable that any other direction for the magnetization averaged over the whole sample, see the orange dots in Fig.~\ref{fig:JP_generic_Du}.

The high field frequency shift, $\Delta f_{hf}$, which is of relevance for extracting anisotropy constants from experiment \cite{gross_dynamic_2016, gross_meso_2021}, first increases with truncation, but later decreases again, see Fig.~\ref{fig:JP_generic_Du} (c). This is owed to the loss of magnetic material for increasing truncation as seen in Fig.~\ref{fig:JP_generic_Du} (b).

In contrast to the SW model the micromagnetic simulations are able to correctly reproduce the experimental curve shape of $\Delta f$ for the hard axis orientation, see Fig.~\ref{fig:JP_generic_triple} (d).
\begin{figure}[t]
  \includegraphics[width=7cm]{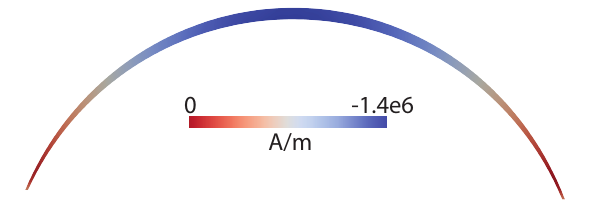}
  \caption{Cut through the geometry of the JP at the position of the $xz$ plane showing the $x$ component of the demagnetizing field within the magnetic layer for $\mathbf{H}\parallel\mathbf{\hat{x}}$ and $\mu_0H = \SI{20}{\tesla}$.}
\label{fig:demag_field}
\end{figure}
To understand the reason for this, the $x$ component of the demagnetizing field $H_{demag,x}$ within the magnetic layer is visualized for a cut through the geometry in Fig.~\ref{fig:demag_field} at high applied field in $x$ direction. It shows a gradual change of the demagnetizing field magnitude with $z$ position, which is a good measure of the preferred orientation of a magnetic moment (top part with $H_{demag,x}\approx 0$ prefers $x$ orientation, opposing to bottom part with maximum $H_{demag,x}$, which needs maximum external field to be aligned in $x$ direction).
This shows, that magnetic moments in proximity to the pole need the smallest field magnitude to be aligned with a field in $x$ direction. The required field magnitude gradually increases the closer a magnetic moment is situated to the equator. This explains the gradual change of $\Delta f$ with increasing field in this orientation, opposing to what is evident in the SW model, where all magnetic moments rotate in unison.
\section{Generic simulations of truncated spherical halfshells with and without exchange bias}
\label{sec:generic}
\begin{figure*}[t]
  \includegraphics[width=17cm]{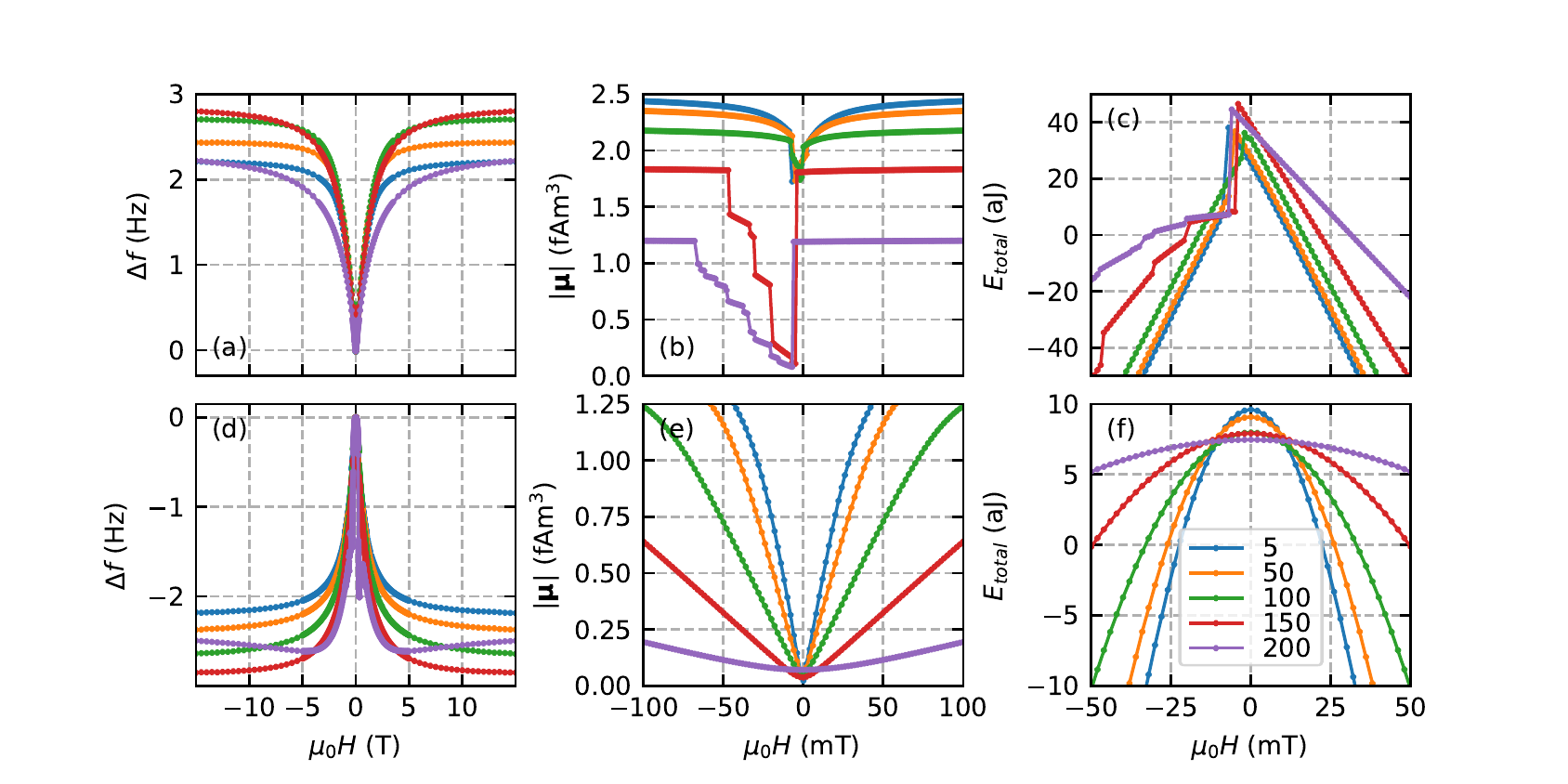}
  \caption{Frequency shift $\Delta f$, total magnetic moment $\mu$ and total energy $E_{total}$ for JPs with different truncation as indicated in the legend in \si{\nano\meter}. Top row: External field applied in the easy plane ($\mathbf{H}\parallel\mathbf{\hat{x}}$). Bottom row: External field applied in the hard direction ($\mathbf{H}\parallel\mathbf{\hat{z}}$).}
\label{fig:JP_generic_triple}
\end{figure*}

The speedup of simulations due to a reduced size of the JPs, as
defined in section \ref{sec:canti}, allows us to analyze the influence
of simulation parameters such as the truncation on the magnetic
hysteresis. In Fig.~\ref{fig:JP_generic_triple} a set of data for JPs
with different truncations is shown. Besides the differences in high
field asymptotes as already discussed in section \ref{sec:shape}, the
truncation also significantly changes the curvature of $\Delta f(H)$,
see Figs.~\ref{fig:JP_generic_triple} (a) and (d). By adjusting the
truncation, this allows to fit the curvature of a given experimental
$\Delta f(H)$ in the simulation.

The magnetic state evolves very much as discussed in section
\ref{sec:progression} for field alignment in both easy
($\mathbf{H}\parallel\mathbf{\hat{x}}$) and hard
($\mathbf{H}\parallel\mathbf{\hat{z}}$) orientation. Yet, especially
for the former we observe some distinct differences for a truncation
of $d>\SI{100}{\nano\meter}$: Rather than through the formation of an
S-state close to remanence, magnetic reversal occurs through a vortex
state, that only appears for small reverse fields. This manifests
itself e.g.\ in the magnitude of the total magnetic moment
$|\mathbf{\mu}|$, which is significantly less for the vortex state as
compared to the S-state, as shown in Fig.~\ref{fig:JP_generic_triple}
(b).

For the external magnetic field perpendicular to the easy plane, we find that magnetic reversal takes place via vortex formation for all values of the truncation. Consequently, the total magnetic moment at remanence is always small, as seen in Figs.~\ref{fig:JP_generic_triple}~(b) and (e).

There is a big difference between the magnetic moments at remanence depending on through which progression remanence has been reached. Hence, in order to find the most stable state, it is instructive to compare the energies of the different progressions as shown in Figs.~\ref{fig:JP_generic_triple} (c) and (f). A state with significant $|\mathbf{\mu}|$ at remanence always seems to possess higher energy than the vortex state.  Compare, for example, the energy at remanence for the JP with \SI{200}{\nano\meter} truncation for easy (no vortex) and hard (vortex state) field alignment. The former has about \SI{40}{\atto\joule}, while the latter is more favorable with only \SI{7}{\atto\joule}.\\

As discussed in the main text, exchange bias, which forces magnetic
moments into a certain direction, can avoid a global vortex state at
remanence. We simulate the magnetic hysteresis of reduced-size JPs
with global unidirectional anisotropy and vary its strength
$K_{eb}$. As an example, we pick a JP with a cut of
$\SI{100}{\nano\meter}$, and match the orientation of the
unidirectional anisotropy to the parallel field direction, which gives
the maximum effect in the DCM signal.

\begin{figure*}[t]
  \includegraphics[width=17.2cm]{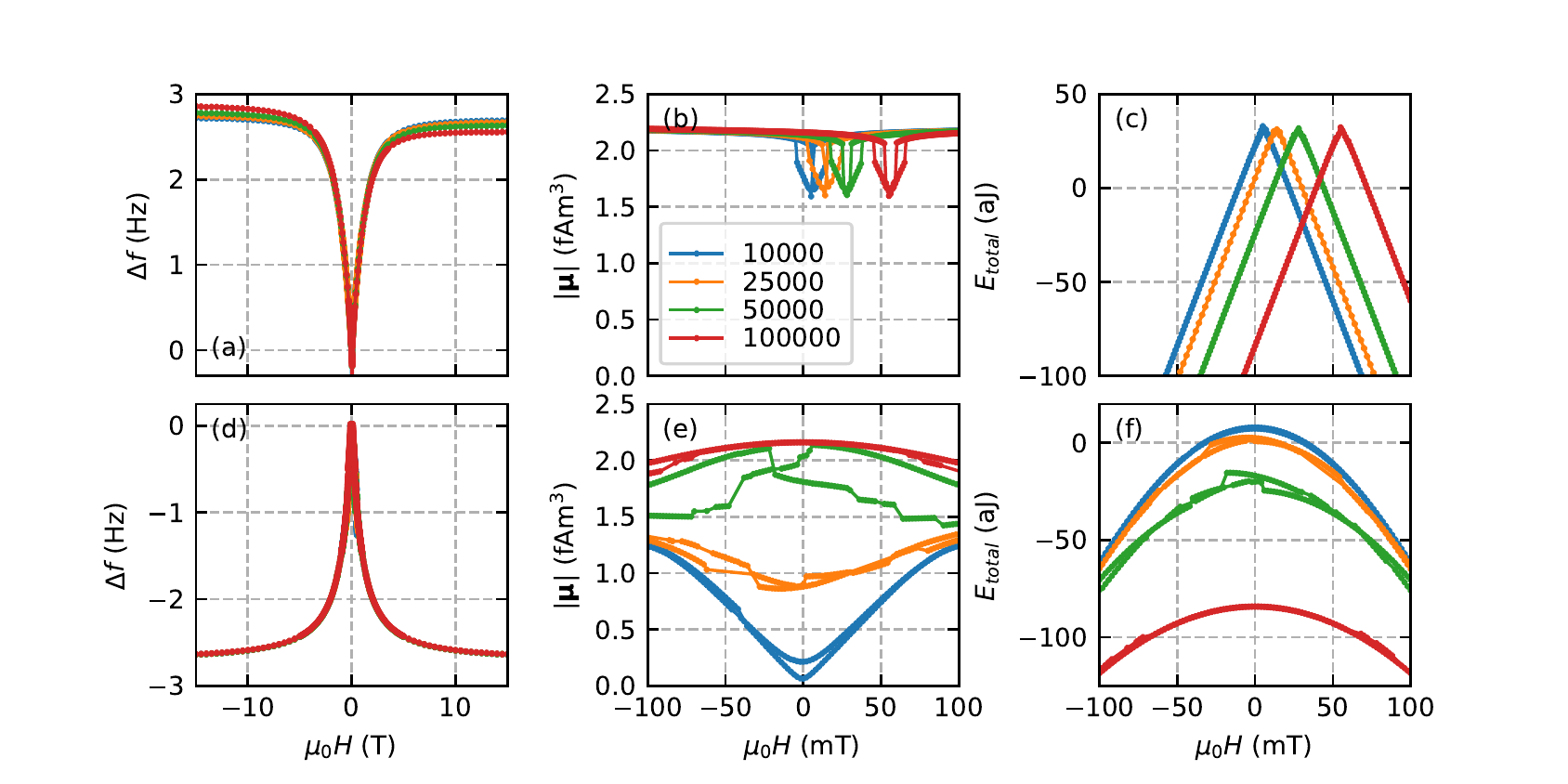}
  \caption{Frequency shift $\Delta f$, total magnetic moment $\mu$ and total energy $E_{total}$ for JPs with \SI{100}{\nano\meter} truncation and different strengths of the unidirectional anisotropy as indicated in the legend in \si{\joule/\meter\cubic}. Top row: External field applied in the easy plane ($\mathbf{H}\parallel\mathbf{\hat{x}}$). Bottom row: External field applied in the hard direction ($\mathbf{H}\parallel\mathbf{\hat{z}}$).}
\label{fig:JP_generic_eb_triple}
\end{figure*}

In Fig.~\ref{fig:JP_generic_eb_triple}, we plot the same set of data
for the JP with exchange bias as for the purely ferromagnetic JPs in
Fig.~\ref{fig:JP_generic_triple}. The frequency shift for high fields
for $\mathbf{H}\parallel\mathbf{\hat{x}}$ develops an increasing
asymmetry for increasing strength $K_{u}$ of the unidirectional
anisotropy, as shown in Fig.~\ref{fig:JP_generic_eb_triple} (a). In
turn, for perpendicular alignment this asymmetry is not evident, as
shown in Fig.~\ref{fig:JP_generic_eb_triple}~(d). Hence, the
experimentally observed asymmetry in the asymptotes can serve as good
indicator for the strength of the undirectional anisotropy, given that
its orientation is known.
The main influence of the unidirectional anisotropy on the progression
of the magnetic state with external field for easy field alignment is
to shift magnetic reversal in magnetic field magnitude. This can e.g.\
be seen in $|\mathbf{\mu}|$, see Fig.~\ref{fig:JP_generic_eb_triple}
(b). Magnetic reversal takes place around zero field for
$K_{eb}=\SI{1}{\kilo\joule/\meter\cubic}$, and is shifted to happen
around \SI{50}{\milli\tesla} for
$K_{eb}=\SI{100}{\kilo\joule/\meter\cubic}$. The dominant state during
reversal remains an onion state for all values of $K_{eb}$.
For hard alignment, a vortex appears in the JP for all values of
$K_{eb}$, just as described in section \ref{sec:ebjp}. The vortex is
centered in the JP for larger field magnitudes, but moves to the side
of the JP, if the field is reduced. Depending on $K_{eb}$, the vortex
moves only very little (small values for $K_{eb}$), moves
significantly to the side of the JP (intermediate $K_{eb}$), or even
escapes the JP through the equatorial line (large values of
$K_{eb}$). If the latter happens, the vortex reenters the JP in
reverse field from the other side of the JP and then moves back to the
center for increasing reverse field. 
The larger the $K_{eb}$, the larger is the total magnetic moment
$\mu$ at remanence, for $K_{eb}=\SI{100}{\kilo\joule/\meter\cubic}$ we
find almost full saturation in direction of the anisotropy vector, see
Fig.~\ref{fig:JP_generic_eb_triple} (e).

A clear trend is observable, if we consider the energy that sets which
remanent state is more likely over long time scales, as shown in
Figs.~\ref{fig:JP_generic_eb_triple} (c) and (f): The difference in energy between the remanent states gets
smaller for increasing $K_{eb}$, and hence diminishes the relevance of
the chosen orientation of an applied field to magnetize the JP. The
trend for the total magnetic moment is, irrespective of the which
orientation for the external field is chosen, to be larger in
magnitude if $K_{eb}$ is increased. 
%
%
\section{DCM in the high-field limit with unidirectional anisotropy}
\label{sec:Dfhfunidir}
The high-field limit in DCM is reached for
$|H| \gg |\sum_i K_i/(\mu_0 M_s)|$, where $K_i$ are the different
anisotropy contributions for a given direction. The frequency shift of
the cantilever resonance is determined in this limit by the
competition of the different anisotropy contributions, and can be
calculated analytically~\cite{gross_dynamic_2016,
  gross_meso_2021}. For the presence of unidirectional anisotropy of
strength $K_{eb}$ this is given by:
%
%
\begin{equation}
\begin{split}
&\Delta f_{\text{unidir}} = -\frac{f_0 V K_{eb}}{2k_0 l_e^2}\cdot 
\\& \bigg(\cos\theta_u (\sin\theta_h \sin\theta_{eb} \cos\phi_u \cos\phi_{eb}+\cos\theta_h \cos\theta_{eb})
\\& -\sin\theta_{eb} (\sin\theta_u \cos\theta_h \cos\phi_{eb}+\sin\theta_h \sin\phi_u \sin\phi_{eb})
\\& +\sin\theta_u \sin\theta_h \cos\theta_{eb} \cos\phi_u\bigg)
\end{split}
\label{eq:limit}
\end{equation}
Here, $(\theta_h,\phi_h = 0)$ define the orientation of the external field, $(\theta_u,\phi_u)$ of the axis of the uniaxial shape anisotropy as defined in Ref.~\onlinecite{gross_meso_2021}, and $(\theta_{eb},\phi_{eb})$ of the unidirectional anisotropy vector. The latter is oriented first, and rotated by $(\theta_u,\phi_u)$ in a second step to be consistent with the situation in experiment. Cantilever and magnetic parameters are as defined before.

\begin{figure}[t]
  \includegraphics[width=7cm]{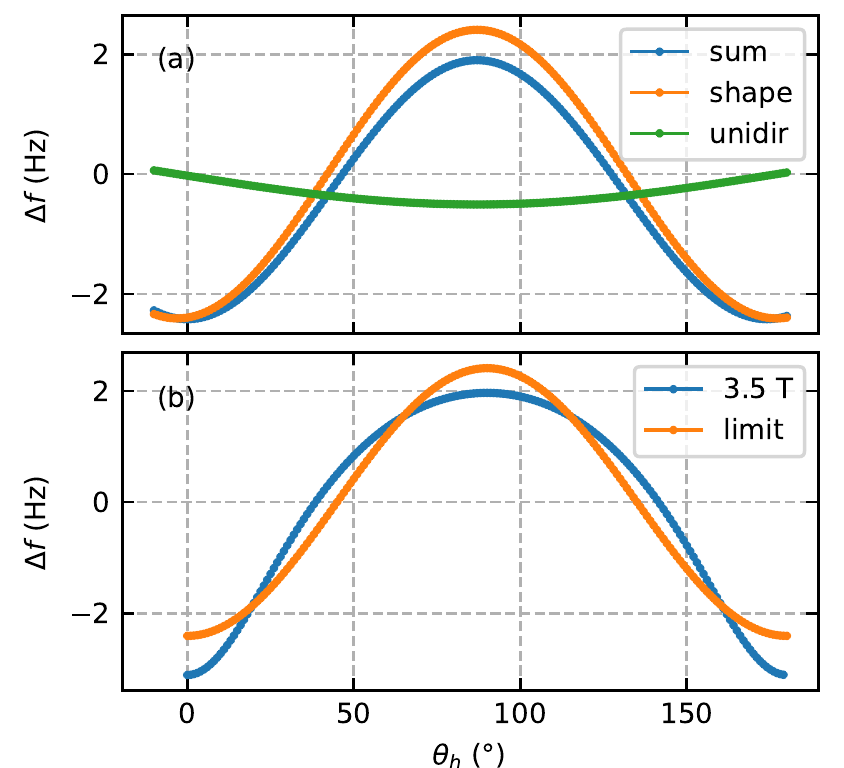}
  \caption{(a) $\Delta f(\theta_h)$ in the high field limit with shape and unidirectional anisotropy. $\Delta f(\theta_h)$ in the SW model for \SI{3.5}{\tesla} applied field magnitude, and in the high field limit from (a).}
\label{fig:hf_unidir}
\end{figure}

We evaluate the high-field limit for the sum of shape and
unidirectional anisotropy for an exemplary situation as it may be
present for the exchange biased JPs, using similar parameters as in
section \ref{sec:shape}. However, we increase $K_{eb}$ significantly
to magnify its effects. The angles are set to be
$(\theta_u,\phi_u)=(\SI{-3}{\degree},\SI{0}{\degree})$ and
$(\theta_{eb},\phi_{eb})=(\SI{-90}{\degree},\SI{0}{\degree})$, while
$\theta_h$ is varied as in experiment. The result is shown in
Fig.~\ref{fig:hf_unidir} (a), together with the individual
contributions from shape and unidirectional anisotropy.
This shows, that the sum of the two contributions may lead to a
periodicity that deviates slightly from \SI{180}{\degree}, which wold be 
given for pure uniaxial shape anisotropy. Furthermore, the
magnitude of maxima and minima may differ significantly. Here we find \SI{1.9}{\hertz} for the maxima and \SI{2.4}{\hertz} for the minima, respectively.
We have indeed observed such large asymmetries in experiment for low
temperatures (not shown here), however, the origin is different, as
will be discussed in the following. Nevertheless, for strong unidirectional
anisotropies these findings should be observable in experiment.

The SW model, as described in section~\ref{sec:SWmodel}, can be used
to calculate $\Delta f(\theta_h)$ for a fixed field magnitude, as done
in experiment for \SI{3.5}{\tesla}. This allows to compare the result
of the SW model with the high field limit as discussed above, see Fig.~\ref{fig:hf_unidir} (b).  The curve of
the high field limit follows a (negative) cosine with $2\theta_h$ in
the argument. In turn, for the SW model at \SI{3.5}{\tesla}, minima
are deeper and maxima are shallower in $\Delta f$, respectively. However, there is no
deviation from the \SI{180}{\degree} periodicity. Further, maxima are
wider than minima, which is a consequence of the fact that positive
and negative asymptotes are approached with a different curvature when
ramping up the external field in the SW model, compare curves in the
last panel in the first column with those in the last panel in the
last column in Fig.~\ref{fig:SWeasyplane}. In experiment this
behavior of the maxima and minima is inverted, which is caused by
the extreme curvature of the magnetic layer JPs, as discussed in
section \ref{sec:shape}.

\bibliographystyle{unsrt}